# Noise Equivalent Counts Based Emission Image Reconstruction Algorithm of Tomographic Gamma Scanning*

WANG Ke(王珂)[1,2], LI Zheng(李政)[1,2; 1)], FENG Wei(冯伟)[1,2], HAN Dong(韩冬)[1,2]

[1]Key Laboratory of Particle & Radiation Imaging (Tsinghua University), Ministry of Education, Beijing 100084, China

[2]Department of Engineering Physics, Tsinghua University, Beijing 100084, China

**Abstract**:

Tomographic Gamma Scanning (TGS) is a technique used to assay the nuclide distribution and radioactivity in nuclear waste drums. Both transmission and emission scans are performed in TGS and the transmission image is used for the attenuation correction in emission reconstructions. The error of the transmission image, which is not considered by the existing reconstruction algorithms, negatively affects the final results. An emission reconstruction method based on Noise Equivalent Counts (NEC) is presented. Noises from the attenuation image are concentrated to the projection data to apply the NEC Maximum-Likelihood Expectation-Maximization algorithm. Experiments are performed to verify the effectiveness of the proposed method.

**Key words**: Tomographic Gamma Scanning; emission image reconstruction; attenuation correction; noise equivalent counts

**PACS**: 29.30.Lw, 28.41.Kw

## 1 Introduction

With the growth of the nuclear energy industry, a lot of nuclear waste drums need to be disposed. Before dealing with the waste drums, it is necessary to obtain the nuclide information to meet the safety requirements. Among the various nondestructive assay methods, Tomographic Gamma Scanning (TGS) technology stands out for its ability to detect the nuclide distribution and radioactivity accurately. The first TGS prototype was developed in the Los Alamos National Laboratory in the early 1990s[1]. And much research has investigated TGS ever since[2-6]. Although commercial TGS devices are available already[7], improvements of TGS are still needed.

TGS needs to scan the waste drum twice, one for transmission measurements with an external isotopic source and another for emission measurements without such a source. The data of the emission measurement is used to reconstruct the emission image which displays the distribution of the radioactive sources in the drum, and the transmission image reconstructed from the transmission measurement data displays the attenuation matrix which is used for the attenuation correction in the emission image reconstruction. As a result, the attenuation matrix plays an important role in the emission reconstruction. However, with the limited scanning time and large voxel size, there are errors in the transmission images which will be transferred to the final results. Currently, the Maximum-Likelihood Expectation-Maximization (ML-EM) algorithm is widely used in the emission image reconstruction of TGS[8], in which the errors of the transmission images are not considered. Some technologies are needed to decrease the influence of the attenuation matrix errors.

An emission reconstruction method is proposed to deal with the problem. The distribution of the

* Supported by the National Natural Science Foundation of China (No. 11175101).

1) Corresponding author. E-mail: lizheng@mail.tsinghua.edu.cn

attenuation coefficients are measured through numerical simulations, and the noise variance of the projection data is then calculated, which doesn't follow the Poisson distribution in ML-EM. To deal with the different noise variance, the noise equivalent counts (NEC) ML-EM algorithm is applied[9,10].

In this paper, the common knowledge of transmission and emission reconstruction is introduced first. Then, the NEC ML-EM algorithm and the computational process of the noise variance of the projection data are presented. Finally, the experimental setup and results are shown to confirm the effective of the methods.

## 2 Methods

The sketch of TGS is shown in Fig. 1. There is only one detector in the TGS system. In the transmission scan, the drum rotates to get projections in different angles. And in each angle, the external isotopic source, the collimator and the detector translate to measure different ray path through the drum. The system acts similar when performing the emission scan except that there's no such an external source.

As mentioned above, the scan is a time consuming job. Hence, the number of measurements is small due to the limited scanning time. Typically, each layer of the drum is divided into 10 × 10 voxels and about 150 projections are needed. The drum rotates 15 or 20 times over 180 degrees, and the source, the detector and the collimator translate 10 times over the diameter of the drum when the aperture of the collimator is the same with the voxel size.

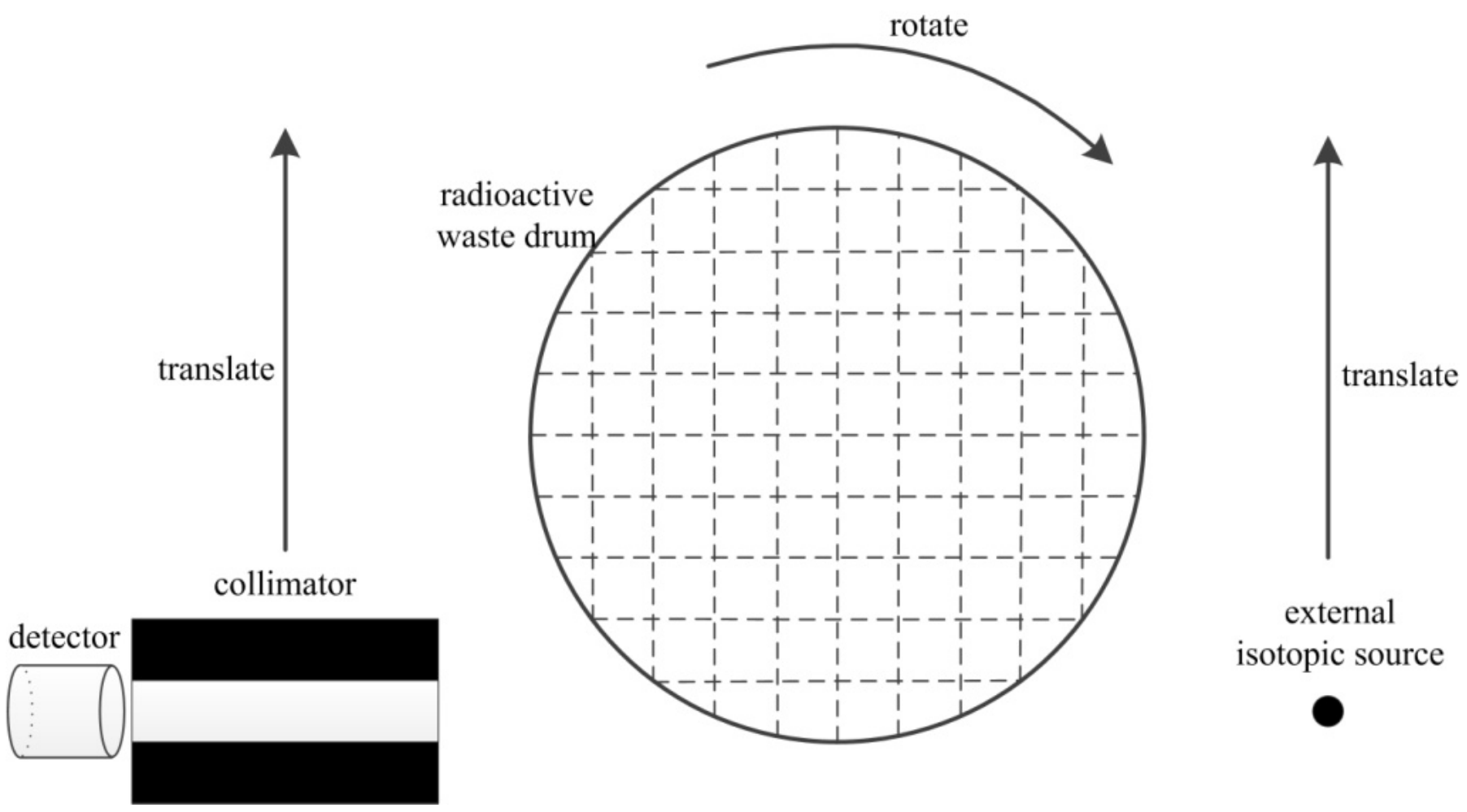

Fig. 1. The sketch of TGS.

### 2.1 Transmission Scan

According to the Beer's law, the transmission scan of TGS can be described as follows:

$$N_i = N_0 \exp(-\sum_j t_{ij}\mu_j) , \quad (1)$$

where $N_0$ is the count rate of a specific energy $E$ which is attenuated by the air only, $N_i$ is the count rate of $i^{th}$ measurement, $t_{ij}$ is the trace length of the $j^{th}$ voxel along the ray from the external isotopic source to the detector, and $\mu_j$ is the attenuation coefficient of the $j^{th}$ voxel to be solved. Each voxel is supposed to be uniform in the transmission scan.

Equation (1) can be converted into a linear form:

$$g_i = \sum_j t_{ij} \mu_j , \tag{2}$$

where $g_i$=-ln($N_i/N_0$). Algebra reconstruction technique (ART) algorithm[1,11] is preferred in the transmission image reconstruction of TGS:

$$\mathbf{f}^{(r+1)} = \mathbf{f}^{(r)} + \frac{g_i - T_i \mathbf{f}^{(r)}}{\left\| T_i \right\|^2} T_i^T , \tag{3}$$

where $\mathbf{f} = \{\mu_j\}$, $\boldsymbol{T}_i = \{t_{ij}\}|_{j=1:J}$, and $r$ is the iteration number.

## 2.2 Emission Scan

The emission scan measures the gamma-ray counts from the nuclear waste drum. The ray sum of a specific energy $E$ in the $i^{th}$ measurement is defined as follows:

$$p_i = \sum_j \varepsilon_{ij} a_{ij} s_j = \sum_j h_{ij} s_j , \tag{4}$$

where $p_i$ is the count rate of the $i^{th}$ measurement, $h_{ij}=\varepsilon_{ij}a_{ij}$, and $\varepsilon_{ij}$ is the detection efficiency if a radioactive source is placed in the center of the $j^{th}$ voxel without any attenuation, $s_j$ is the activity of the $j^{th}$ voxel to be solved. $a^{ij}$ is the attenuation correction factor described as follows:

$$a_{ij} = \prod_m \exp(-t_{ijm} \mu_m) = \exp(-\sum_m t_{ijm} \mu_m) , \tag{5}$$

where $\mu_m$ is the attenuation coefficient of the $m^{th}$ attenuation voxel which can be obtained from the transmission scan results, $t_{ijm}$ is the the length in voxel $m$ along the ray connecting the emission voxel $j$ and the detector. Equation (4) can be converted into a matrix form:

$$P = HS , \tag{6}$$

where $\boldsymbol{P}=\{p_i\}$, $\boldsymbol{H}=\{h_{ij}\}$ and $\boldsymbol{S}=\{s_j\}$. $\boldsymbol{H}$ is the system matrix of the emission measurement. The emission image is the solution of Equation (6).

## 2.3 NEC ML-EM

Among the various image reconstruction algorithms, $\boldsymbol{H}$ is supposed to be accurate which is true in most conditions. However, in Equation (6), $\boldsymbol{H}$ is related to the attenuation image which is from the transmission image reconstruction. The error of the transmission image will be transferred to $\boldsymbol{H}$. Until now, the errors of $\boldsymbol{H}$ are not concerned in the emission reconstruction of TGS. An algorithm based on ML-EM and concerning the error of $\boldsymbol{H}$ will be derived.

$p_i$ is usually supposed to follow the Poisson distribution in emission image reconstructions. Because of the existence of the error of the attenuation process, we assume that the noise variance of $p_i$ is

$$\sigma^2(p_i) = k_i \overline{p_i} , \tag{7}$$

where $\overline{p_i}$ is the expectation of $p_i$ and $\sigma^2(p_i)$ means the variance of $p_i$. $k_i$ is from the bias of attenuation image.

The original ML-EM algorithm is

$$s_j^{(r+1)} = \frac{s_j^{(r)}}{\sum_i H_{ij}} \sum_i H_{ij} \frac{p_i}{\sum_l H_{il} s_l^{(r)}} \tag{8}$$

where $r$ is the iteration number. And according to the NEC model, when the noise variance of $p_i$ can be expressed as Equation (7), we can modify the ML-EM algorithm[9,10] as:

$$s_j^{(r+1)} = \frac{s_j^{(r)}}{\sum_i \frac{1}{k_i} H_{ij}} \sum_i \frac{H_{ij}}{k_i} \frac{p_i}{\sum_l H_{il} s_l^{(r)}} . \tag{9}$$

To obtain the estimation of $k_i$ in Equation (7), the error distribution of the attenuation coefficient $\mu_m$ is needed. The Monte Carlo Replicate (MCR) method[12] is used to estimate the distribution. The numerical simulation model is shown in Fig. 2. The model has 10 × 10 voxels which is the same as the transmission image resolution in TGS. The value of the 4 × 4 pixels in the center ranges from 0.02 to 0.32. The gamma-ray count without any attenuation is $10^4$. Transmission projections are applied to the model and then Poisson noises are used to smear the projection data. Then, the transmission image is reconstructed by performing ART algorithm. At last, repeat the above process for 10,000 times and record the value of each voxel.

The distribution of the reconstructed voxel values are shown in Fig. 3. Distributions of the random 4 non-zero voxels are displayed as examples. Fig. 3 shows that the attenuation coefficients follow the Gaussian distribution. The relationship between the means and the variances are shown in Fig. 4. The variance of the attenuation coefficient trends to get larger as the attenuation coefficient increases. We assume a linear relationship between the variance and the attenuation coefficient:

$$\sigma^2(\mu) = w\mu , \tag{10}$$

where $w$ is the parameter describing the linear equation. And when $\mu$ is zero, $\sigma^2(\mu)=0$ fits the reality better. So, the intercept of the line is set to 0 manually.

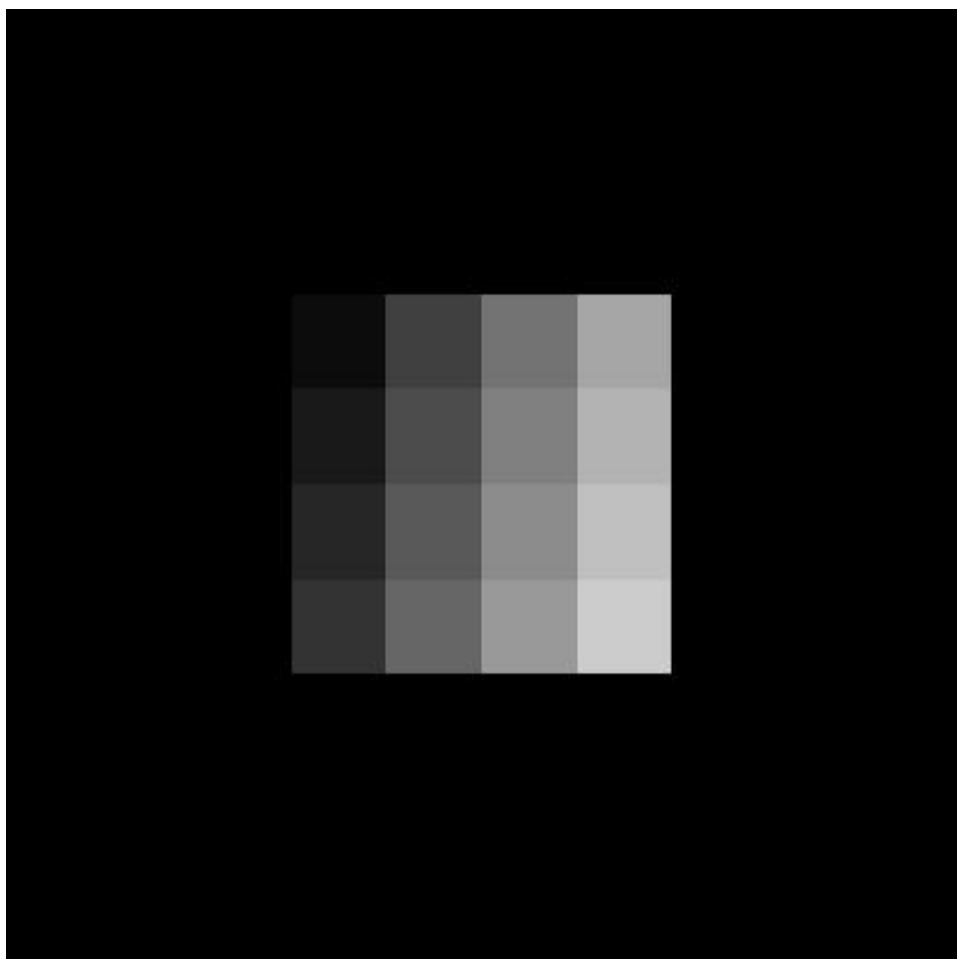

Fig. 2. Numerical simulation model, 10 × 10 voxels. The gray scale window is [0 0.4].

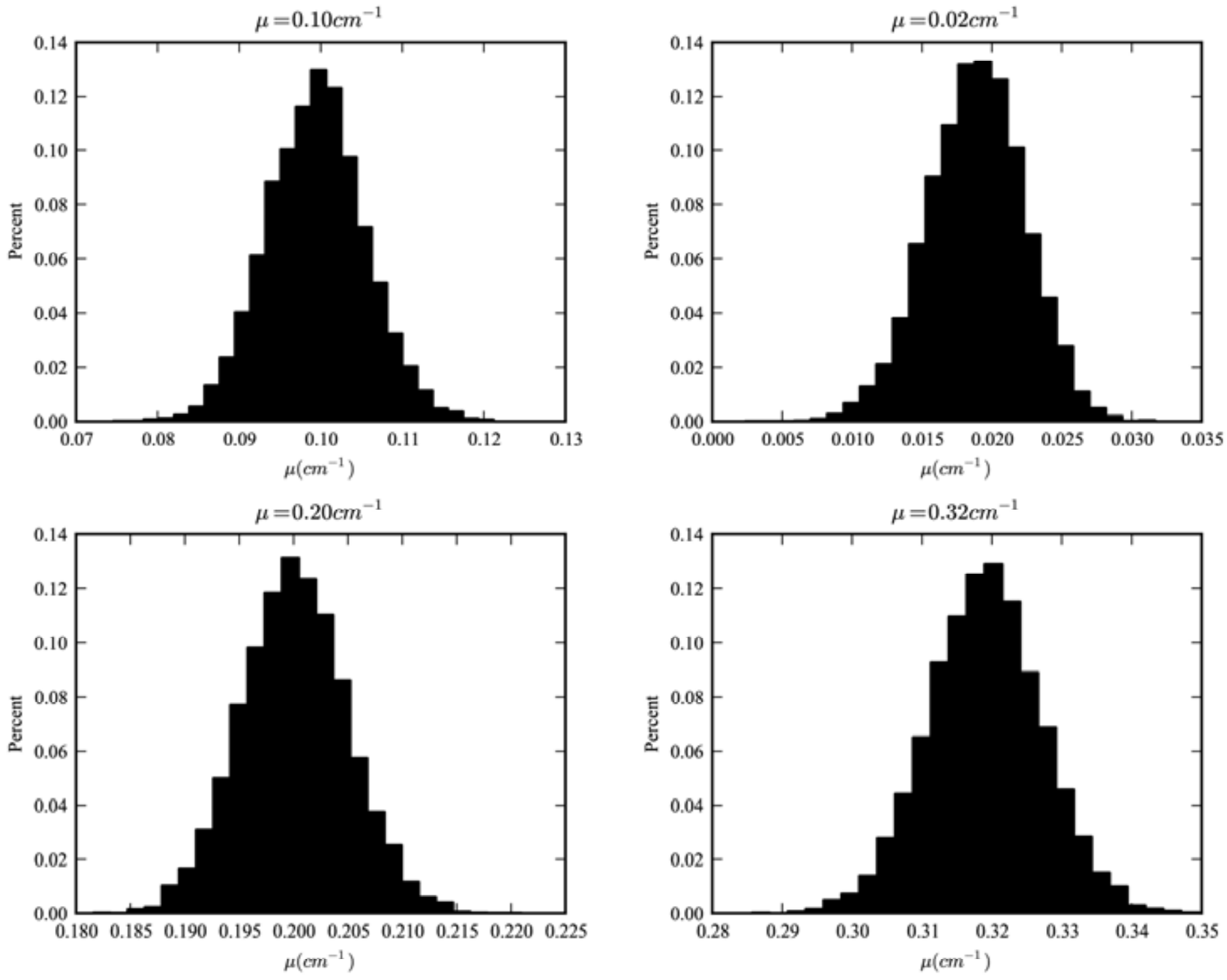

Fig. 3. Distribution of the attenuation coefficients.

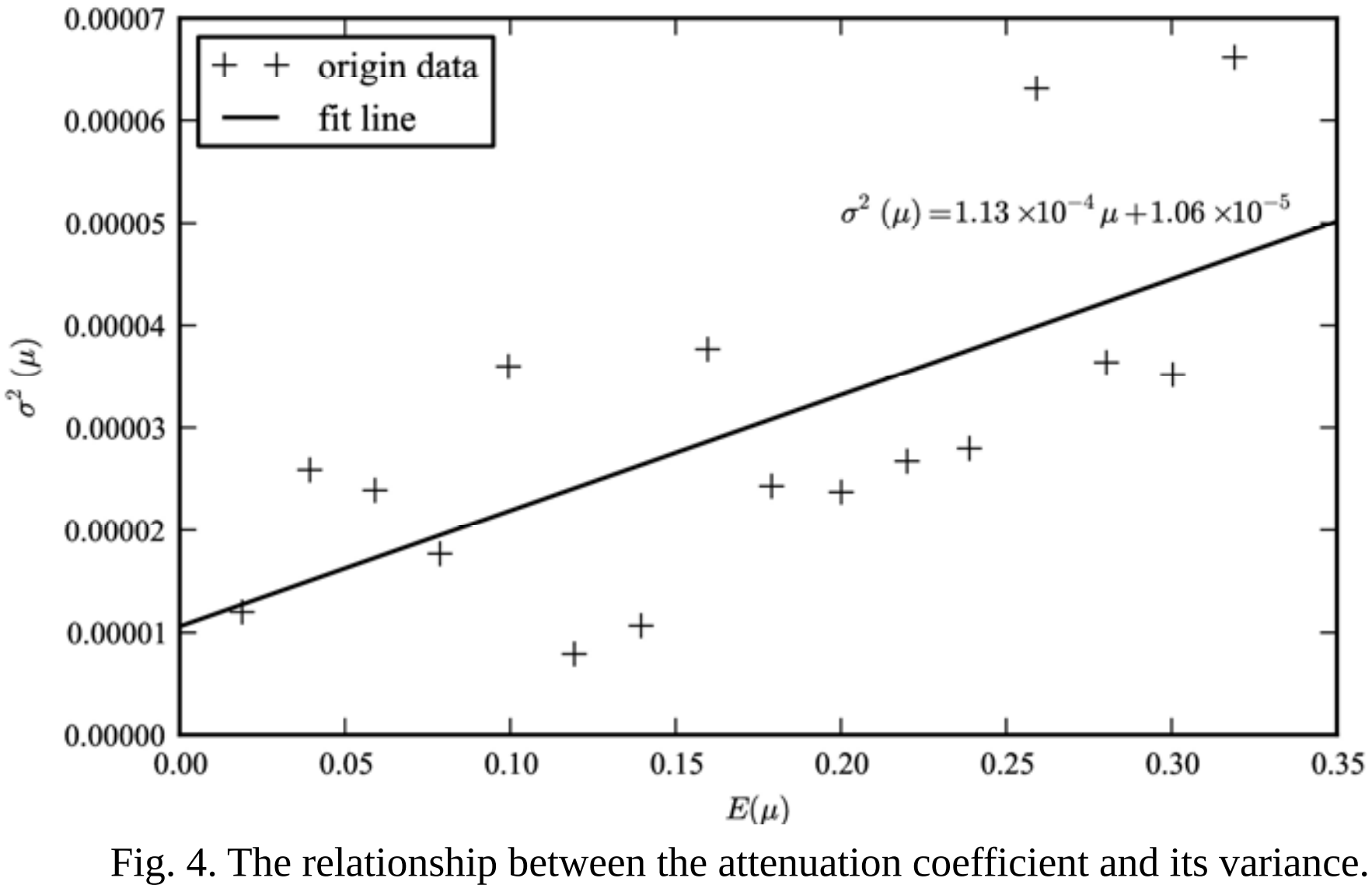

Fig. 4. The relationship between the attenuation coefficient and its variance.

According to Equation (5), $a_{ij}$ follows the lognormal distribution as $\mu_m$ follows the Gaussian distribution. The variance of $a_{ij}$ is

$$\sigma^2(a_{ij}) = (e^{\sigma^2(v_{ij})} - 1)e^{-2v_{ij} + \sigma^2(v_{ij})}, \tag{11}$$

where $v_{ij}=\sum_m t_{ijm}\mu_m$ , $\sigma^2(v_{ij})=w\sum_m t_{ijm}{}^2\mu_m$ . Then, the variance of $p_i$ can be expressed:

$$\begin{aligned}\sigma^2(p_i)&=\sum_j \sigma^2(a_{ij}\varepsilon_{ij}\hat{s}_j)\\&=\sum_j \sigma^2(a_{ij})[\sigma^2(\varepsilon_{ij}\hat{s}_j)+E(\varepsilon_{ij}\hat{s}_j)^2]+\sigma^2(\varepsilon_{ij}\hat{s}_j)E(a_{ij})^2\end{aligned}, \tag{12}$$

where $\hat{s}_j$ is an estimation of $s_j$. $\hat{s}_j$ can be obtained by reconstructing the emission image with ML-EM. Suppose that $\varepsilon_{ij}\hat{s}_j$ follows the Poisson distribution whose variance and expectation are equal, then Equation (12) becomes

$$\sigma^2(p_i)=\sum_j \sigma^2(a_{ij})[\varepsilon_{ij}\hat{s}_j+(\varepsilon_{ij}\hat{s}_j)^2]+a_{ij}{}^2\varepsilon_{ij}\hat{s}_j , \tag{13}$$

where $\varepsilon_{ij}\hat{s}_j$ is substituted for $E(\varepsilon_{ij}\hat{s}_j)$, and $a_{ij}$ is substituted for $E(a_{ij})$.

Comparing Equation (7) and Equation (13), we can get

$$k_i=\frac{\sum_j \sigma^2(a_{ij})[\varepsilon_{ij}\hat{s}_j+(\varepsilon_{ij}\hat{s}_j)^2]+a_{ij}{}^2\varepsilon_{ij}\hat{s}_j}{p_i}. \tag{14}$$

Equation (9) and Equation (14) are the NEC ML-EM algorithm in TGS.

## 3 Experiments and results

The voxel size of a commercial TGS device is approximately 6 cm, and each layer is divided into 10 × 10 voxels. Limited to the experiment conditions, the prototype used here is smaller than the real TGS. The maximum diameter of the object that can be scanned is 20 cm. The object is also divided into 10 × 10 voxels, which means the voxel size of each voxel is 2 cm. The length of the collimator is 15 cm and the side length of the square aperture is 2 cm. The distance between the center of rotation and the front end of the detector is 26.5 cm, and the extern isotopic source is 20 cm apart from the center of rotation. For simplicity, only one layer is scanned in the experiments.

The two models used in the experiments are shown in Fig. 5. Fig. 5 (a) stands for model 1 and Fig. 5 (b) stands for model 2. Four kinds of materials are used in both models, which are graphite, polyethylene, iron and aluminum. All the materials in the objects are cylindrical. A $^{137}$Cs source is placed in the center of the objects as the emission source. The external transmission source is also a $^{137}$Cs. Both transmission scan and emission scan are performed to the two models. The total activity of the results from different emission reconstruction methods will be compared. In the emission reconstruction, the detection efficiency matrix is from a Monte Carlo simulation program based on Geant4.

To verify the effectiveness of our algorithm, another special emission scan is performed where no attenuation material exists in the object and the $^{137}$Cs emission source is at the same place. To exclude the interference of the error from the detection efficiency matrix, the result of this emission scan will be treated as the real activity of the $^{137}$Cs source.

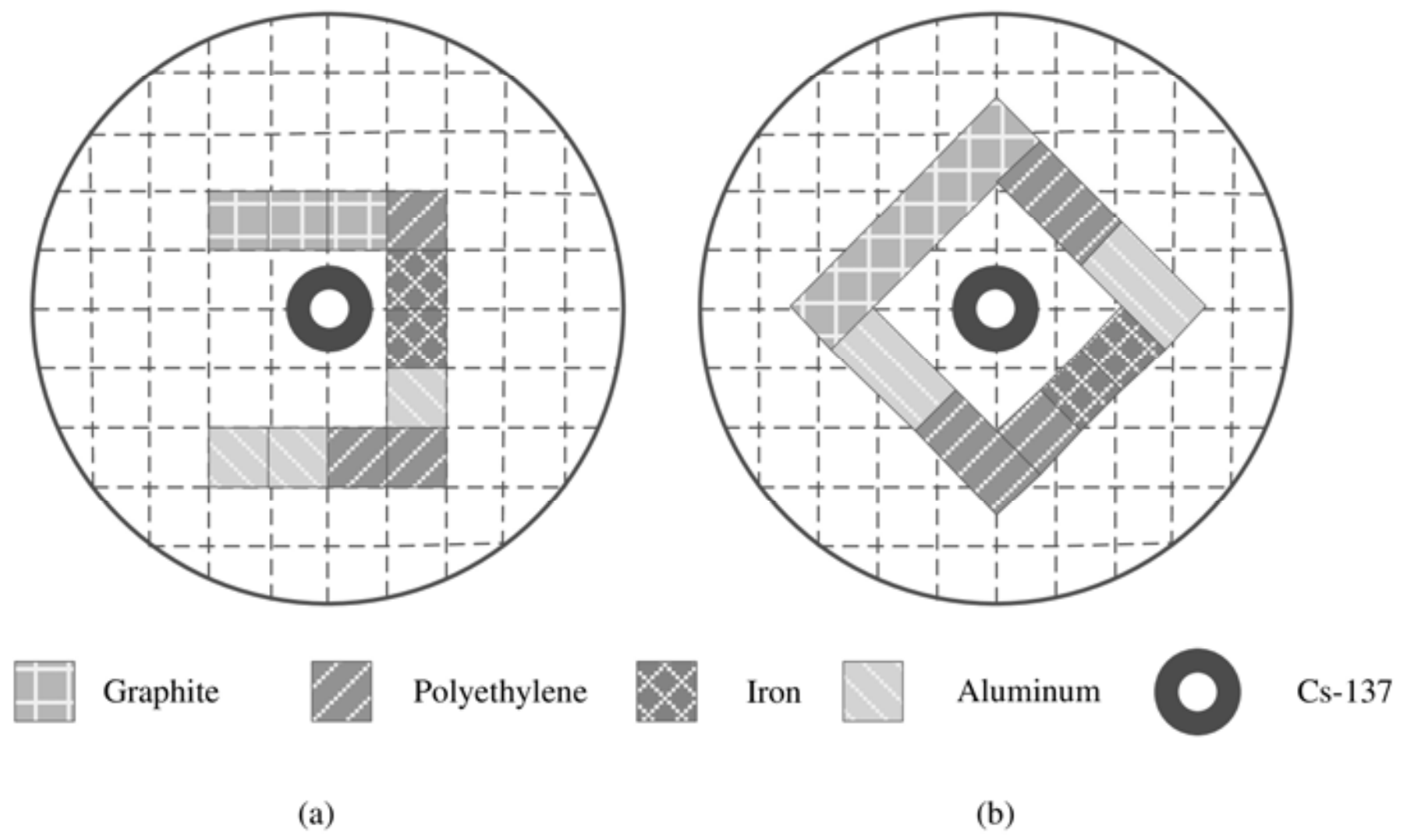

Fig. 5. Experimental models. (a) Model 1. (b) Model 2.

The transmission images are shown in Fig. 6, (a) for model 1 and (b) for model 2. ART algorithm is used for the transmission image reconstruction. Both transmission images are 10 × 10 voxels. The emission results are shown in Table 1. In Table 1, the true activity, from the special non-attenuation emission scan, is 4.42 μCi. The activity from NEC ML-EM reconstruction is more accurate than that from original ML-EM reconstruction in both model 1 and model 2. The parameter $w$ in Equation (10) is set to $1.13 \times 10^{-4}$ as shown in Fig. 4.

The essence of the NEC ML-EM is that different weight factors are used for different projection data in the back-projection process. And the weight factors mainly come from the noises of the attenuation correction factors. Those more accurate measurement data will certainly have heavier weight.

Compare model 1 with model 2, the left side of the source in model 1 is not surrounded with materials which results in more attenuation imbalance than model 2. And in Table 1, the NEC ML-EM improves more on model 1 than on model 2. In other words, NEC ML-EM has better improvement on the uneven distribution drums.

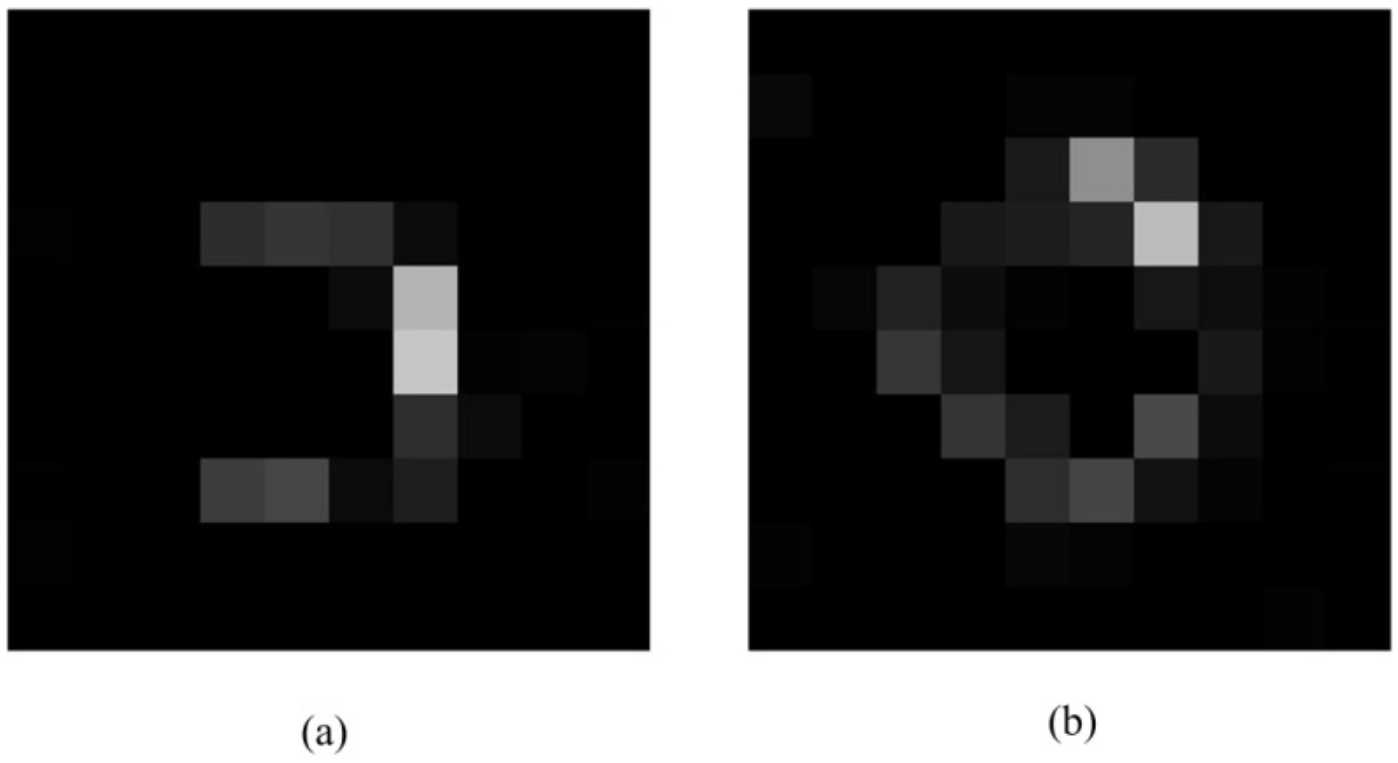

Fig. 6. Transmission images, 10 × 10 voxels.
The gray scale window is [0 0.6]. (a) Model 1. (b) Model 2.

Table 1 Activity and error

| Model Number | ML-EM | | NEC ML-EM | |
|---|---|---|---|---|
| | Activity/μCi | Error | Activity/μCi | Error/% |
| Model 1 | 4.01 | -9.28% | 4.28 | -3.17% |
| Model 2 | 4.20 | -4.98% | 4.27 | -3.39% |

## 4 Conclusion

This study proposes a method based on the NEC ML-EM to reduce the influence of the error of the attenuation correction factors in the emission image reconstruction. The attenuation coefficients are found follow the Gaussian distribution through MCR method, and the noise variance of emission projection data is calculated based on this. Experiments are performed to verify the effectiveness. The experimental results show that the NEC ML-EM method leads to better total isotopic activity. And the method works better on those drums with uneven distributions.

## 5 Acknowledgement

This work is supported by the National Natural Science Foundation of China (No. 11175101).